\documentclass[pra,twocolumn,showpacs]{revtex4}

\usepackage{amssymb}
\usepackage{amsfonts}
\usepackage{amsmath}
\usepackage{epsfig}

\newcommand{\Xstate}{\mbox{X$^1\Sigma^+$}}
\newcommand{\astate}{\mbox{a$^3\Sigma^+$}}
\newcommand{\rcm}{\mbox{cm$^{-1}$}}
\newcommand{\wn}{$\mathrm{cm}^{-1}$}

\begin{document}

\title{The coupling of the X$^{1}\Sigma ^{+}$ and a$^{3}\Sigma ^{+}$ states of KRb}

\author{A.~Pashov}
\affiliation{Department of Physics, Sofia University, 5 James Bourchier boulevard, 1164 Sofia, Bulgaria}

\author{O.~Docenko}
\author{M.~Tamanis}
\author{R.~Ferber}
\affiliation{Department of Physics and Institute of Atomic Physics and Spectroscopy, University of
Latvia, 19 Rainis boulevard, Riga LV-1586, Latvia}

\author{H.~Kn\"ockel}
\author{E.~Tiemann}
\affiliation{Institut f\"ur Quantenoptik, Leibniz Universit\"at Hannover, Welfengarten 1, 30167 Hannover,
Germany}


\begin{abstract}
A comprehensive study of the electronic states at the 4s+5s
asymptote in KRb is presented. Abundant spectroscopic data on the
\astate\ state were collected by Fourier-transform spectroscopy
which allow to determine an accurate experimental potential energy curve up to
14.8 \AA\ . The existing data set (C. Amiot et al.
J. Chem. Phys. 112, 7068 (2000)) on the ground state \Xstate\ was
extended by several additional levels lying close to the atomic
asymptote. In a coupled channels fitting routine complete
molecular potentials for both electronic states were fitted. Along
with the line frequencies of the molecular transitions, recently
published positions of Feshbach resonances in $^{40}$K and
$^{87}$Rb mixtures (F. Ferlaino et al. Phys. Rev. A 74, 039903
(2006)) were included in the fit. This makes the derived potential
curves capable for an accurate description of observed cold
collision features so far. Predictions of scattering lengths and
Feshbach resonances in other isotopic combinations are reported.
\end{abstract}


\maketitle


\section{Introduction}

\label{intro}

Along with NaK and NaRb, the KRb molecule is one of the most
extensively studied among the mixed alkalis. The ground
state \Xstate, the lowest singlet states and some of the triplet states
were studied by various spectroscopic techniques. It is therefore
surprising that the triplet state \astate\ was not characterized
experimentally so far.

In recent years, understanding interactions between different
alkali-metal species in their singlet and triplet ground states of the pairs
became important due to the experiments in mixed atomic traps. This stimulated investigations of the molecular states as well.
Using molecular spectroscopy, several ground state asymptotes were
examined: NaRb \cite{Pashov:05}, NaCs \cite{Docenko:06}, LiCs
\cite{Staanum:07}. In the present study we applied the same technique for
KRb. This molecule is particularly interesting since several
groups are operating two-species traps, cold KRb molecules have
been already formed, positions of several Feshbach resonances in
$^{40}$K$^{87}$Rb have been measured \cite{Ferlaino, Ferlaino:07, Ospelkaus} and new ones are still being searched
for. The study of the (4s)K+(5s)Rb asymptote has therefore several
strong motivations: (i) to obtain complete and accurate potentials for the \Xstate\ and
\astate\ states, to be able to model cold collisions; (ii) to prove
of the validity of the isotopic scaling rules, i.e. applying the
potentials derived for the main isotopic combination
$^{39}$K$^{85}$Rb to reproduce the experimental observations
(including the Feshbach resonances) in other isotopes of KRb;
(iii) to predict new Feshbach resonances that can be later
verified and precisely measured experimentally with much reduced effort for the search procedure.

Along with the theoretical calculations of the long range
dispersion coefficients \cite{Bussery:87, Marinescu:99,
Derevianko:01a, Porsev:03} there is only one theoretical work yet devoted to the
study of the electronic structure of KRb \cite{Rousseau:00}.

The ground state \Xstate of KRb was first studied for v$'' \le 44$ at high
resolution by Ross and coworkers \cite{Ross:90}. A significant
step towards the asymptote was done in Ref. \cite{Amiot3:00},
where an accurate potential energy curve for v$''$ up to 87 was
derived. In both investigations the laser induced fluorescence to the KRb ground state
was recorded by exciting the A$^1\Sigma^+$ state by radiation of a Ti:Sapphire
laser. The B(1)$^1\Pi$ state was studied in
Refs.~\cite{Leininger:97,Kasahara:99,Amiot2:00}. Numerous
perturbations were observed due to mixing with the nearby lying
triplet states. Therefore in order to study the \astate\ state it was
expected to follow straightforward the strategy which was successfully
applied fot the investigation of the same states in NaRb \cite{Pashov:05}
and NaCs \cite{Docenko:06}, i.e. to use the perturbations in the
B(1)$^1\Pi$ state as a window to the triplet manifold.

\section{Experiment}
\label{Exper}

\begin{figure}
   \centering
\epsfig{file=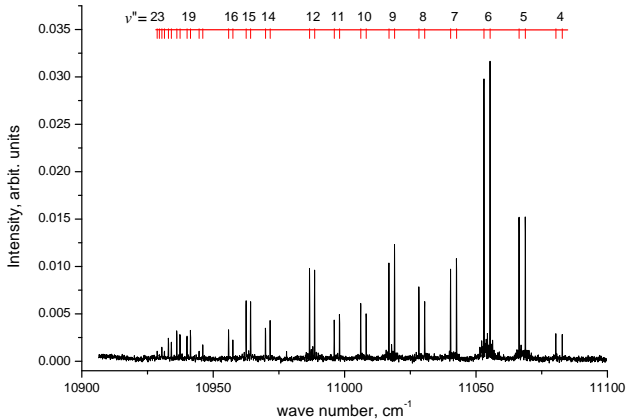,width=0.99\linewidth}
  \caption{(color online) A fragment of a triplet progression excited at 15059.018 \rcm\ and
originating from ($v',J'=36$) in the B - X system. The doublets of the progression are
formed by transitions to levels with $N''= 36$ and 38 of the
\astate\ state. For strong lines (e.g. v$''$=6) collisional
satellites are also visible. }\label{trips}
\end{figure}

The KRb molecules were produced in a stainless steel heat-pipe
similar to that developed for NaRb \cite{Pashov:05} and NaCs
\cite{Docenko:06}. Metallic K and Rb  were mixed in an approximate
mass ratio 1:1. The heat-pipe was operated typically at a
temperature of 550 K with 2-5 mbar of Ar as buffer gas. The Laser
Induced Fluorescence (LIF) was excited with narrow band single
mode lasers and resolved with a Fourier Transform Spectrometer
(FTS) Bruker 120 HR.

For excitation of the B(1)$^1\Pi$ state we used a Coherent 599
linear laser with DCM dye delivering a power of about 70 mW. In
the region between 15000 \rcm\ and 15300 \rcm\ we found a lot of
transitions in the B-X band system. Along with them, transitions
in the similar system of K$_2$ were often seen with comparable
intensity. Going to higher frequencies the transitions in K$_2$
dominated those in KRb. It is worth noting here, that when we
tested a heat-pipe filled only with potassium, the fluorescence
due to K$_2$ at similar experimental conditions (temperature,
laser frequency, laser power) was at least a factor of 3 stronger
than in the KRb heat-pipe. Obviously in case of mixtures the
concentration of molecules cannot be derived simply from the
saturated vapour pressure of the correspondent pure metals as
often stated in spectroscopic papers. This is confirmed also by
the fact that Rb$_2$ fluorescence was not observed in the studied
spectral region, although Rb$_2$ was reported to absorb there
\cite{Amiot:93,Amiot:97,Caldwell:80}.

\begin{figure*}
  \centering
  \centering
\epsfig{file=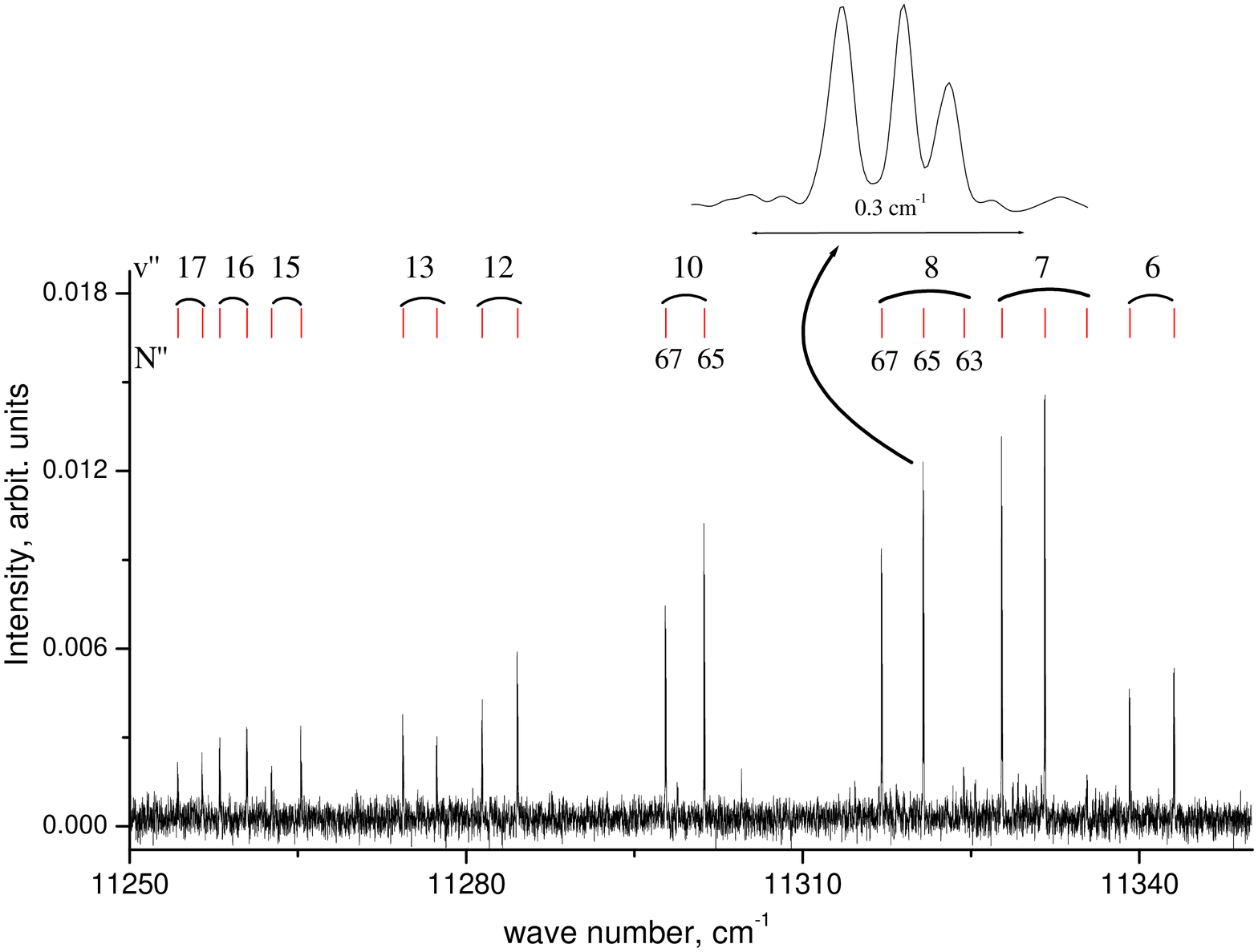,width=0.50\linewidth}
\epsfig{file=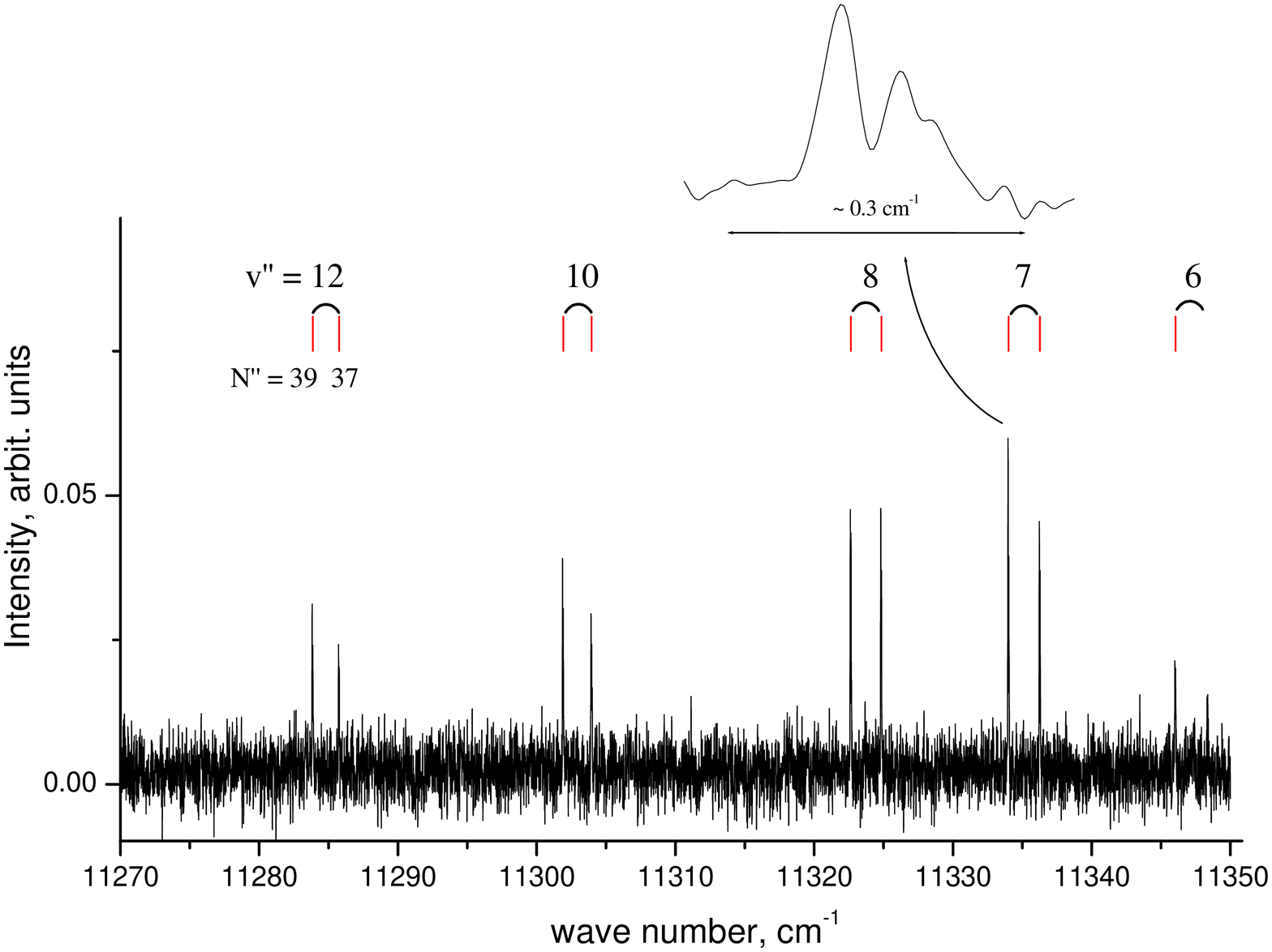,width=0.48\linewidth}
  \caption{(color online) Examples of ``usual'' and ``unusual'' HFS within the
progressions to the \astate\ state. On the left side the
progression starts from an upper state level with $J'$=65. The
doublets are formed by transitions to levels with $N''=67$ and 65,
except those with $v''= 7$ and 8 where a weak transition to
$N'=63$ also appears. On the right side a transition from $(v',
J'=39)$ is shown. Both insets give examples of hyperfine structure, details see text.  }\label{HFS}
\end{figure*}

\begin{figure}
  \centering
  \centering
\epsfig{file=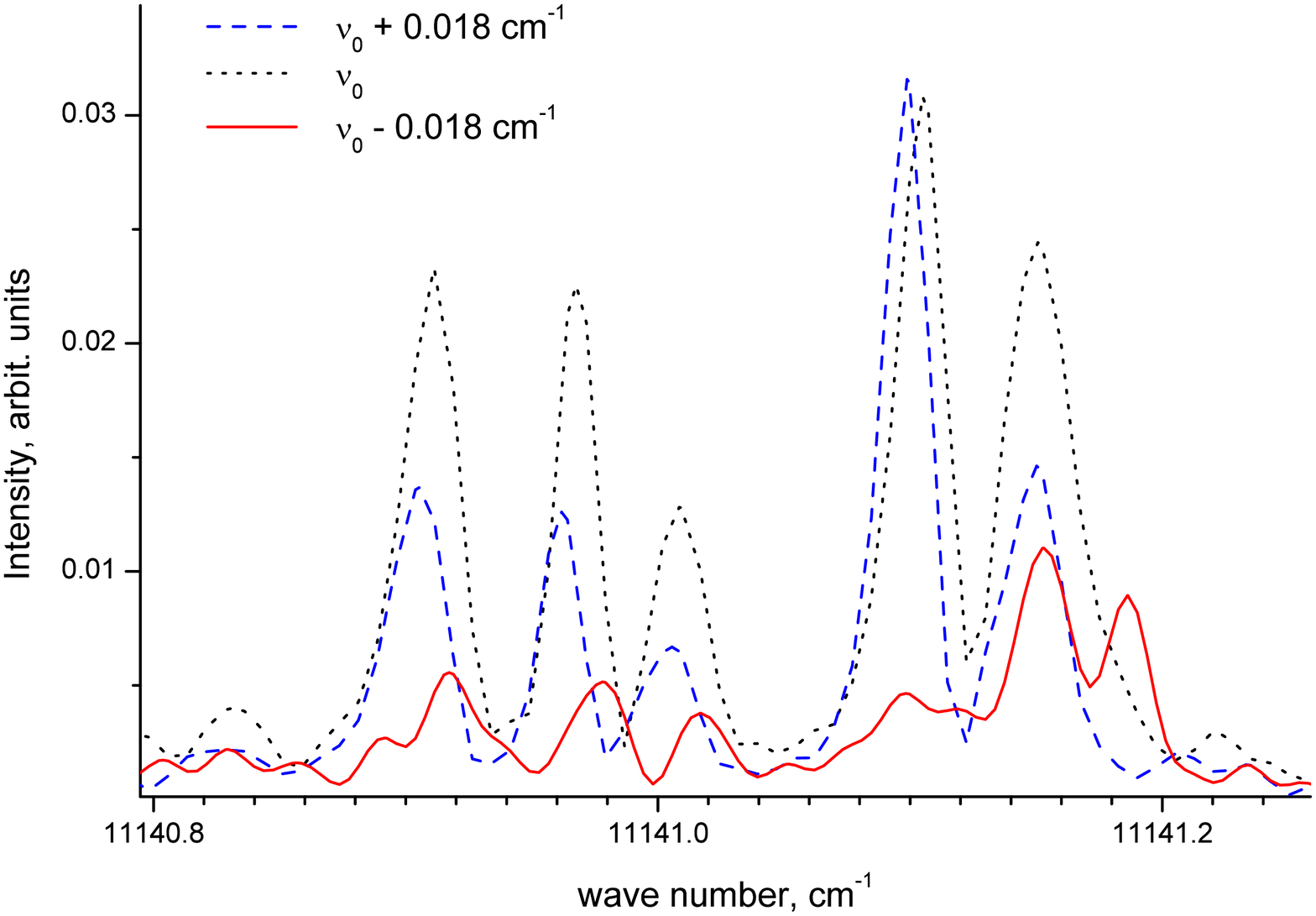,width=0.99\linewidth}
  \caption{(color online) Two hyperfine groups of closely spaced transitions to \astate\ state
levels are shown for three different frequencies of the exciting
laser. These transitions belong to different progressions, i.e.
they originate from different excited levels ($v',J'=9$ and
$v',J'=12$). It is clearly seen that while the structure of the
left transition (to the ($v''=6,N''=9$) \astate\ state level)
remains almost unchanged and only the intensity varies, the
overall pattern of the right transition (to the ($v''=6,N''=11$)
level) changes; for the blue and red detuned excitation
frequencies, different components of the hyperfine triplet are
missing.}\label{HFS2}
\end{figure}

The fluorescence from the B state led usually back to the ground
\Xstate\ state. However when exciting a perturbed level with triplet
admixture, fluorescence down to the triplet state \astate\ was
seen as well. It appeared in the region of 11000-12000 \rcm\, well
separated from the B-X fluorescence. An example is given in 
Fig.~\ref{trips},where the fluorescence progression extends 
close to the dissociation asymptote, which can be concluded 
from the large decrease of the vibrational spacing below 10950 \rcm . 
When resolved by the FTS with
a resolution of typically 0.03 \rcm\, most lines to the \astate\
state show the expected hyperfine structure (HFS) similar to that
observed in Ref.~\cite{Pashov:05,Docenko:06}. In those cases it
could be explained by the Fermi contact interaction
\cite{Kasahara:96} using the atomic HFS parameters of Rb and K
\cite{Arimondo}. However, contrary to the NaRb and NaCs cases
\cite{Pashov:05,Docenko:06}, there was also a large number of
progressions all lines of which show unusual HFS (see
Fig.~\ref{HFS}). This we shall discuss in the following section~\ref{analysis}.

The ground state \Xstate\ was studied in Ref.~\cite{Amiot3:00} up
to v$''$=87.  The classical outer turning point of the last observed
level is about 11.3 \AA. This value differs from
that given in the title of Ref.~\cite{Amiot3:00}, namely 10~\AA , since the latter
corresponds to the turning point of the level  ($v''$=87,
$J''$=0), but in fact the authors observed that vibrational level for
$J''$=51 which shifts the classical turning point to a larger
internuclear distance. This distance, however is still too small
in order to study the coupling between the \astate\ and 
\Xstate\ states due to the HFS mixing. Therefore, we revisited the
\Xstate\ state asymptote in order to extend the experimental
information closer to the dissociation limit. For this purpose we excited the \mbox{A$^1\Sigma^+$} - \Xstate system 
using a Ti:Sapphire laser (from Tekhnoscan) pumped by a 10 W cw
frequency doubled Nd:YAG laser (Coherent Verdi 10) and repeated as first steps 
measurements from \cite{Amiot3:00} for internal calibration check.
Then we extended the experimental data set by recording
progressions from the same upper vibrational level but with
different J$'$. It is worth noting that some of the lines of the
progression reported in Ref.~\cite{Amiot3:00} which go to
v$_X''$=87 turned out to be overlapped with stronger lines from
different progressions. Exciting other upper vibrational levels that
would give fluorescence to high lying ground state levels was
not applicable due to the presence of strong K$_2$ fluorescence in the
same spectral region.

We searched also for transitions to the \astate\ state due to the
so called long-range-changeover which provided precious
information in the case of NaRb \cite{Pashov:05} and LiCs
\cite{Staanum:07}. By exciting a B state level close to the atomic
asymptote (in those cases (n$^2$S)+(m$^2$P)) we actually approach
the range where Hund's case (c) starts to be more appropriate
than Hund's case (a) (due to the large fine structure
splitting of the lowest p state in Rb and Cs). As a consequence, a 
decay from such excited levels to both \Xstate\  ($\Omega=0^+$) and
\astate\ ($\Omega=1$ component) states is possible, which cannot
be explained within the Hund's case (a) or (b) models either due
to selection rules or due to vanishing Franck-Condon factors for the case of 
local perturbations. The case of KRb, however, turned out to
be more similar to that of NaCs \cite{Docenko:06}, where the
long-range changeover appeared rather fragmentary and provided
only few transitions to near asymptotic levels in the \Xstate\ and
\astate\ states.

\section{Analysis}
\label{analysis}

Our previous experience \cite{Pashov:05,Docenko:06,Staanum:07} has
shown, that the states \astate\ in the heteronuclear alkali-metal
dimers can be modeled within the experimental resolution in the
Born-Oppenheimer approximation by a single potential curve while
the hyperfine splitting is well described within our resolution
with the atomic hyperfine parameters. This is possible if the hyperfine
structure does not depend on the internuclear separation. Only close to the
atomic asymptote it is necessary to take into account the mixing
with the ground states \Xstate .

The usual appearance of the hyperfine splitting of the \astate\
levels is shown in Fig.~\ref{HFS} (inset of left graph). The splitting into
three components is due to the interaction of the electrons with
the nuclear spin of the Rb atoms. Each of these components is
further split due to the interaction with the K nuclear spin. This
interaction is small, contrary to the case of Na in NaRb for example
\cite{Pashov:05}, and is below our experimental
resolution. 

In the case of KRb we observed for the first time
several progressions where the HFS differs from the expected one
(see Fig.~\ref{HFS}). We made special efforts to study these
``unusual'' features within the possibilities of our experimental
setup. We found out that within a given progression the hyperfine splitting is always
independent of the rotational and vibrational quantum numbers of
the \astate\ (both for the usual and the unusual cases). At the
same time transitions to one given \astate\ level coming from
different excited levels can show different structure. Therefore, we
concluded that it should be the structure of the upper state
levels which is responsible for the deviations from the expected
HFS. This idea was confirmed by the observation that the
unusual HFS changes its appearance when the excitation
laser is detuned within the Doppler profile of the molecular transition. In
Fig.~\ref{HFS2} we show results of such an experiment. The
laser excites simultaneously two transitions, one
(($v''=0,J''=9$)$\rightarrow$($v',J'=9$)) gives rise to a
progression with usual HFS ($N''=7,9,11$ are involved, \cite{hundb}) and the
second (($v''=0,J''=11$)$\rightarrow$($v',J'=12$), the assignment
of $J'$ is not certain, which is unimportant for the analysis of the ground states)
 - with unusual one ($N''=11,13$ are
involved). We chose a portion of the fluorescence spectrum where two lines,
each belonging to one of the progressions, appear closely
together. Then we set the laser frequency within the Doppler
profile of the transitions at the three position (center and $\pm$ 0.018 \rcm and observed the change of the HFS of
these lines. It is clearly seen in Fig.~\ref{HFS2}  that while the
structure of the line with N$''$=9 remains almost unchanged and
varies only in intensity, the overall pattern of the line with
N$''$=11 changes. For the blue and red detuned excitation
frequencies, different components of the hyperfine triplet are
missing. Moreover, from the relative changes of the intensities of
the lines from one laser frequency to another we can expect that
the effective widths of the excitation transitions are different.
The transition producing the progression with unusual HFS could be
broader, which is probably due to HFS of the upper state level,
being comparable with the Doppler width of the transition. This
hypothesis seems to be reasonable if we compare it with the 
HFS of the (2)$^3\Sigma^+$ state in NaRb observed by
Matsubara {\it et al.} in Ref.~\cite{Matsubara:93}. So, presently,
we understand our observations in the following way: The HFS of
the excited complex of singlet and triplet states in KRb is in some
cases larger than the Doppler width of the transitions. The
laser with given frequency excites a selected set of hyperfine
components which in turn does not give fluorescence to the full
multiplet of the \astate\ levels. Therefore, the
observed structure of the spectral lines reflects the HFS both of
the lower and the upper electronic states.

As a center frequency for a given hyperfine multiplet we chose the
frequency of the central component of the HFS triplet (see the
left part of Fig.~\ref{HFS}) which is normally also the strongest. For the potential fit we used these
frequencies and included the term energy of the upper level as a
free parameter, common for all frequencies forming one
progression. This selection is possible, because in a given progression the
appearance of the HFS turned out to be independent of the
vibrational and rotational quantum numbers within the experimental
resolution. In case of progression with unusual HFS, we took a
characteristic feature of the hyper fine group of lines (usually
the strongest peak). In this way we were able to reproduce
correctly the vibrational and rotational spacings of the \astate within a
progression. We did not make efforts to reproduce the unusual HFS
of the transitions, because our present understanding of the
problem indicates that we need to include also the HFS of the
upper state, the study of which is outside the scope of this work
and would ask for a different experimental setup to get sufficient data for a convincing analysis.

\subsection{The \astate\ state}

Establishing the rotational numbering of the triplet progressions
was straightforward because almost always they were accompanied by
the corresponding progression to the ground \Xstate\ state. For
finding the proper vibrational numbering we first decided to rely
on the possibility of scaling the molecular constants of one
electronic state with the ratio of the reduced masses of different isotopomers.
Experimentally, progressions belonging to the molecule with the
less abundant isotope $^{87}$Rb were easy to distinguish by the
broader HFS compared to $^{85}$Rb. In our initial analysis we had
several progressions in $^{39}$K$^{85}$Rb and one in
$^{39}$K$^{87}$Rb. The range of N$''$ \cite{hundb} was from 31 to 68 and v$''$ spanning a range of 15 vibrational
quanta. Several fits of Dunham-type coefficients with different
vibrational assignment including both isotopomers indicated the most probable one which we
took for further analysis of the experimental data and potential
energy curve (PEC) construction.

Once the initial vibrational numbering was suggested, we fitted a
potential curve for the \astate\ state using a point-wise
representation \cite{ipaasen} and used it to assign new
vibrational progressions. Periodically, as the size of the
experimental data set significantly increased, the PEC was refined.
We noticed, however, that for larger N$''$ there was a growing
systematic disagreement between the observed and the calculated
transition frequencies, which reached about 0.05 \rcm\ for the
highest N$''$. After all experimental data were assigned, this
problem remained. Similarly to the
\astate\ state in NaRb \cite{Pashov:05}, we can exclude the
possibility for significant spin-rotational and second order
spin-orbit effects in the \astate\ state of KRb. Therefore, we searched for a 
connection between these systematic disagreements and the presence
of unusual HFS of the transitions to the \astate\ state levels - a
phenomenon observed only in KRb up to now. Careful examination of the
spectra and collection of new experimental observations proved
that such a connection is not indicated by our data set.
Therefore, we were forced to revise the initially established
vibrational numbering, since our experience from the \astate\
state in NaRb had shown that a wrong vibrational numbering could lead
to similar systematic disagreements for high rotational quantum
numbers. We tried several hypothesis of shifting v$''$ by up to 2
vibrational quanta and finally for one of them we got an excellent
description of the whole data set within the experimental
uncertainty.

Here we want to direct special attention to this problem, because it
is common practice to establish vibrational numbering by using the
mass relations between the isotopes. Our experience in NaRb and
KRb \astate\ states demonstrates now, that given a limited set of levels with assinged
rotational and vibrational quantum numbers, it is still possible to
achieve an almost satisfactory quality of the fit assuming an
incorrect vibrational assignment. In both molecules, discussed as
examples here, we were able to construct PECs that fit the
experimental data up to N$''$=70, but turned out to be
systematically inadequate when going to higher N$''$. It is
important to note also that in both cases the lowest vibrational
levels ($v''=0-2$ for KRb) were not observed experimentally thus
missing in the data field.

\begin{figure}
  \centering
\epsfig{file=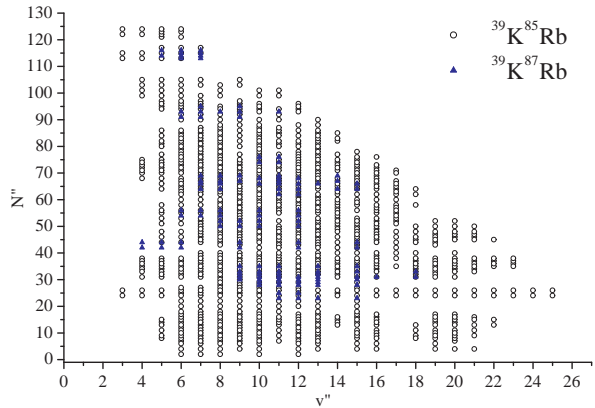,width=0.99\linewidth}
  \caption{(color online) The range of vibrational and rotational quantum numbers of
the energy levels of the \astate\ state, observed in the present study.}
\label{dataset}
\end{figure}

Finally, the analysis of our experimental data led to the
assignment of 2400 transition frequencies to about 1100
rovibrational levels in the \astate\ state in KRb.  In
Fig.~\ref{dataset} the range of the observed triplet state energy
levels is presented. The full list of the transition frequencies is
given in Table I of the supplementary material.

\subsection{The \Xstate\ state}

After the assignment of the levels of the \astate\ state  and the derivation of a preliminary potential for it, we made
first coupled channel calculations including the two states
correlated to the 4s+5s atomic asymptote and the hyperfine
interaction \cite{Pashov:05}. The purpose was to determine the
region where the \Xstate\ and \astate\ states could be treated
as uncoupled, and to use this region for fitting the PEC for the
\astate\ state. For the \Xstate\ we took the potential given in
Ref.~\cite{Amiot3:00}, extended it by the dispersion coefficients
from Refs.~\cite{Derevianko:01a, Porsev:03}. These calculations indicated that the mixing between
the two states is expected to be strong around $v_X''=86$ and
exactly there we found the largest disagreement between the
results of the single channel fit and the experimental
observation. This disagreement, however could not be fully
explained even after the coupling was taken into account. Therefore,
we decided to reinvestigate experimentally the highest levels of the ground state
in KRb, trying to collect higher vibrational levels with a wider
range of J$''$.

\begin{figure}
  \centering
  \centering
\epsfig{file=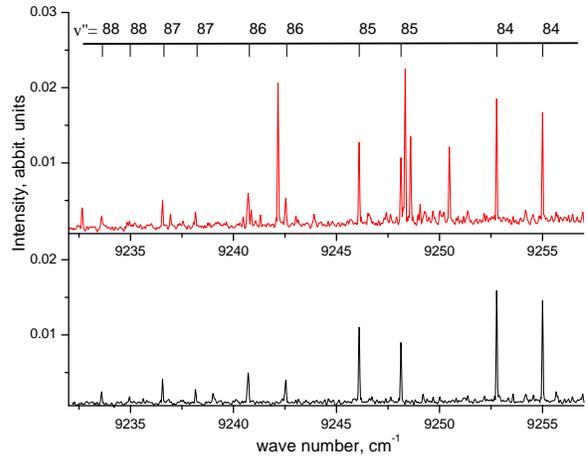,width=0.99\linewidth}
  \caption{(color online) A portion of progressions from the same upper state level (v$'$, J$'$=51),
  but excited from different lower \Xstate\ state levels: (v$''$=9, J$''$=50) for the upper trace and
  (v$''$=8, J$''$=52) for the lower trace.}\label{v88}
\end{figure}

Due to better
signal-to-noise ratio in our experiment we were able to extend the
highest observed vibrational quantum number to v$''$=88. The
corresponding levels were observed for J$''= 51-55$. The assignment of the observed transitions was based on the analysis of Ref.~\cite{Amiot3:00}. As already
mentioned in Section~\ref{Exper} the weak KRb progressions close to the
asymptote were overlapped by stronger K$_2$ transitions. Apparently, this
led to an incorrect determination of the transition frequency to
the (v$''$=86, J$''$=52) level in Ref.~\cite{Amiot3:00} by about
0.02 \rcm. In order to increase the confidence in our assignment
and avoid similar situations we excited the same upper level from
different \Xstate\ levels. Thus, KRb lines of interest should
be the same, but overlapping KRb or K$_2$ lines should be shifted
or disappear. As an example, we show in Fig.~\ref{v88} 
progressions starting from (v$'$, J$'$=51), excited from
(v$''$=8, J$''$=50) for the upper trace and (v$''$=9, J$''$=52) for the lower trace. The same progression
is shown also in Fig.~1 of Ref.~\cite{Amiot3:00}, but there excited from
(v$''$=9, J$''$=52). By a simple inspection it is immediately clear what 
lines appear in both traces and are thus the desired KRb lines. Overall, the present experiment brings 1200 additional lines to the data set of the \Xstate\ state.

In order to make a reliable connection in the energy scale between
the \Xstate\ and the \astate\ states we added to the ground state
data set also levels of progressions (about 600 transitions) that were
recorded simultaneously with progressions to the triplet state and
originating both from a common upper state level. In this way we add in total about 1800 new transition frequencies to the primary data
set obtained in Ref.~\cite{Amiot3:00}.

\section{Potentials}
\label{pot}

The full description of the experimental data needs a model
including adiabatic PECs of the two electronic states and
parameters that describe the coupling between them due to the
hyperfine interaction. Similarly to the approach adopted in
Refs.~\cite{Pashov:05, Docenko:06} we first constructed the
potentials using single channel fits and then we performed coupled
channels calculations and determined the shifts due to the
interaction between the states. As a next step, we corrected the
experimental data with the calculated corrections, obtaining in
this way ``adiabatic'' data sets, i.e. data sets that would have
been observed if the states would not couple. Afterwards the
whole procedure was repeated until the results of the
coupled channels calculations agreed with the experimental
observations.

For the potential construction we first used the pointwise
representation. In the initial stages of fitting it has the
advantage to be very flexible, achieving a convergence in only few
iterations. In this way the potential curve was refined following
the growth of the experimental data set, several hypothesis of the
vibrational numbering in the \astate\ state were tested and also
some wrong line assignments were identified. For the coupled
channels fit the pointwise potentials were transformed into
analytic form \cite{Samuelis:00} with which the number of free parameters is reduced by almost a factor of two.

We split the representation of the potentials into three regions:
the repulsive wall (R$<$R$_{inn}$), the asymptotic region (R$>$R$_{out}$),
and the intermediate region in between.
The analytic form of each potential in the intermediate range is described by a finite power expansion
with a nonlinear variable function $\xi$ of internuclear separation R:

\begin{equation}
\label{xv}
\xi(R)=\frac{R - R_m}{R + b\,R_m}
\end{equation}
\begin{equation}
\label{uanal}
\mbox{U}_{\mathrm {IR}}(R)=\sum_{i=0}^{n}a_i\,\xi(R)^i
\end{equation}

\noindent where \{a$_i$\} are fitting parameters and $b$  and $R_m$
are chosen during the transform process from the spline representation to the analytic one, 
$R_m$ is close to the value of the equilibrium separation. The potential is 
continuously extrapolated for R $ < \mbox{R}_{inn}$ with:

\begin{equation}\label{rep}
\mbox{U}_{\mathrm {SR}}(R)= A + B/R^{6}
\end{equation}

\noindent by adjusting the $A$ and $B$ parameters.

For large internuclear distances (R$ > \mbox{R}_{out}$ )
we adopted the standard long range form of molecular potentials:

\begin{equation}
\label{lrexp}
  U_{\mathrm {LR}}(R)=U_{\infty}-C_6/R^6-C_8/R^8-C_{10}/R^{10}\pm E_{\mathrm{exch}}
\end{equation}

\noindent where the exchange contribution is given by

\begin{equation}
\label{exch}
E_{\mathrm{exch}}=A_{\mathrm{ex}} R^\gamma \exp(-\beta R) 
\end{equation}

\noindent and $U_{\infty}$ is the energy of the atomic asymptote (excluding the
hyperfine energies) with respect to the minimum of the \Xstate\
state. It coincides with the dissociation energy of this state,
$D^{\mathrm X}_{\mathrm e}$. The exchange energy is repulsive for
the triplet state (plus sign in (\ref{lrexp})) and attractive for
the singlet state (minus sign). All parameters in
Eqs.~(\ref{lrexp},\ref{exch}) are common for the \Xstate\ and the
\astate\ states.

As a first guess we used the theoretical estimates of long range
parameters published by Derevianko et al. and Porsev et al.
\cite{Derevianko:01a, Porsev:03} for the \Xstate\ and the triplet
\astate\ state. For the exchange energy amplitude we adopted the
estimation from theoretical potentials in Ref.~\cite{Rousseau:00}.
The continuation at $\mbox{R}_{out}$ was obtained smoothly by
variation of $U_{\infty}$  or $a_0$, the latter is preferred because the
dissociation asymptote is taken as common energy reference for
both states and set to zero here.

For the coupled channels calculations the hyperfine interaction is
taken into account as caused by the Fermi contact interaction. The
effective Hamiltonian for the electronic spin $\bold S$ and the nuclear
spin $\bold I$ defines the interaction parameters:

\begin{equation}
\label{hfs}
 H=A_{\mathrm{Rb}}\cdot S_{\mathrm{Rb}}\cdot I_{\mathrm{Rb}} +
 A_{\mathrm{K}}\cdot S_{\mathrm{K}}\cdot I_{\mathrm{K}}
\end{equation}

From the experimental spectra the A$_{\mathrm{K}}$ and
A$_{\mathrm{Rb}}$ constants could be determined applying Hund's
case $b_{\beta S}$ as basis. Generally, the molecular coupling
parameters correspond to the atomic constants A. But for atoms
forming a diatomic molecule the dependence of these parameters on
the bond length and thus on the vibrational and rotational quantum
numbers cannot be excluded (see Ref.~\cite{Knoeckel:04}). Within
the resolution of our experiment we recorded partially resolved
hyperfine structure of the levels of the \astate\ state and since our
data covered a broad range of such levels we checked the
dependence of A$_{\mathrm{K}}$, A$_{\mathrm{Rb}}^{85}$ and
A$_{\mathrm{Rb}}^{87}$ on vibrational and rotational quantum
numbers. As obtained in other cases of mixed alkalis the observed
hyperfine splitting is described within experimental accuracy by
using the atomic parameters as compiled by Arimondo
\cite{Arimondo}. Then, each special case of strong coupling between the
singlet and the triplet levels is well described by the coupled
channels calculations.

With the definitions by equations \ref{xv} to \ref{hfs}, the
iteration process of potential fitting and calculation of the
corrections by singlet-triplet coupling was performed on all
spectroscopic observations of the \Xstate state by
\cite{Amiot3:00} and from the present work and of the \astate state.
It is especially important to note that we obtained a large set of
fluorescence progressions of \Xstate and \astate with a common
excited level, which fixes the relative energy positions of singlet and
triplet level schemes precisely and gives information on the
exchange energy from levels close to the dissociation asymptote.
In total, 12502 transition frequencies were contained in the fit
and 68 free parameters were varied, which gave a dimensionless
standard deviation $\sigma = 0.68$ for both states together. Tables
\ref{tabX} and \ref{taba} present these parameters, but the values actually given there 
are from the final step of fitting described in the following
section including data from cold collision studies of other authors. For the convenience of the reader also the position of the potential minimum in energy (T$_e$) and internuclear separation (R$_e$) is given in both tables.

\begin{table}
\fontsize{8pt}{13pt}\selectfont
\caption{Parameters of the analytic representation of the \Xstate state potential. The energy reference is the dissociation asymptote. Parameters with $^\ast$ are set for continuous extrapolation of the potential. }
\label{tabX}
\begin{tabular*}{1.0\columnwidth}{@{\extracolsep{\fill}}|lr|}
\hline
   \multicolumn{2}{|c|}{$R < R_\mathrm{inn}=$ 3.000 \AA}    \\
\hline
   $A^\ast$ & -0.731001924$\times 10^{4}$ \wn \\
   $B^\ast$ & 0.590300435$\times 10^{6}$  \wn \AA $^{4}$ \\
\hline
   \multicolumn{2}{|c|}{$R_\mathrm{inn} \leq R \leq R_\mathrm{out}=$ 11.000 \AA}    \\
\hline
    $b$ &   $-0.39$              \\
    $R_\mathrm{m}$ & 4.06818602 \AA               \\
    $a_{0}$ &  -4217.814754 \wn\\
    $a_{1}$ & 0.5530165860346669$\times 10^{1}$ \wn\\
    $a_{2}$ & 0.1402690940409950$\times 10^{5}$ \wn\\
    $a_{3}$ & 0.1048318435177453$\times 10^{5}$ \wn\\
    $a_{4}$ & -0.4087971898030580$\times 10^{4}$ \wn\\
    $a_{5}$ & -0.1817659880541463$\times 10^{5}$ \wn\\
    $a_{6}$ & -0.2739889053154796$\times 10^{5}$ \wn\\
    $a_{7}$ & -0.3930730951201573$\times 10^{5}$ \wn\\
    $a_{8}$ & 0.1755013192862486$\times 10^{6}$ \wn\\
    $a_{9}$ & 0.4366867152340260$\times 10^{6}$ \wn\\
   $a_{10}$ & -0.6031147948206772$\times 10^{7}$ \wn\\
   $a_{11}$ & -0.7948028201366105$\times 10^{7}$ \wn\\
   $a_{12}$ & 0.1172293247025888$\times 10^{9}$ \wn\\
   $a_{13}$ & 0.9209455256556711$\times 10^{8}$ \wn\\
   $a_{14}$ & -0.1533597323741656$\times 10^{10}$ \wn\\
   $a_{15}$ & -0.6982526182585652$\times 10^{9}$ \wn\\
   $a_{16}$ & 0.1418207140035536$\times 10^{11}$ \wn\\
   $a_{17}$ & 0.3514113318792982$\times 10^{10}$ \wn\\
   $a_{18}$ & -0.9529625435553510$\times 10^{11}$ \wn\\
   $a_{19}$ & -0.1096758619400216$\times 10^{11}$ \wn\\
   $a_{20}$ & 0.4743826569413903$\times 10^{12}$ \wn\\
   $a_{21}$ & 0.1477139521952726$\times 10^{11}$ \wn\\
   $a_{22}$ & -0.1768149926253122$\times 10^{13}$ \wn\\
   $a_{23}$ & 0.3496665669778944$\times 10^{11}$ \wn\\
   $a_{24}$ & 0.4947896972615874$\times 10^{13}$ \wn\\
   $a_{25}$ & -0.2373506540656795$\times 10^{12}$ \wn\\
   $a_{26}$ & -0.1033913349509970$\times 10^{14}$ \wn\\
   $a_{27}$ & 0.6234843170080333$\times 10^{12}$ \wn\\
   $a_{28}$ & 0.1588626857834081$\times 10^{14}$ \wn\\
   $a_{29}$ & -0.9740574795906707$\times 10^{12}$ \wn\\
   $a_{30}$ & -0.1741861156434701$\times 10^{14}$ \wn\\
   $a_{31}$ & 0.9410306699092036$\times 10^{12}$ \wn\\
   $a_{32}$ & 0.1289614354386887$\times 10^{14}$ \wn\\
   $a_{33}$ & -0.5220986485346664$\times 10^{12}$ \wn\\
   $a_{34}$ & -0.5776754651606419$\times 10^{13}$ \wn\\
   $a_{35}$ & 0.1278431437418345$\times 10^{12}$ \wn\\
   $a_{36}$ & 0.1182411100178171$\times 10^{13}$ \wn\\
\hline
   \multicolumn{2}{|c|}{$R_\mathrm{out} < R$}\\
\hline
  ${U_{\infty}}$ & 0.0 \wn    \\
 ${C_6}$ &    0.2072097$\times 10^{8}$ \wn\AA$^6$      \\
 ${C_{8}}$ &  0.6509487$\times 10^{9}$ \wn\AA$^8$   \\
 ${C_{10}}$ & 0.2575245$\times 10^{11}$ \wn\AA$^{10}$   \\
 ${A_{ex}}$ & 0.1469387$\times 10^{5}$ \wn\AA$^{-\gamma}$   \\
 ${\gamma}$ & 5.25669    \\
 ${\beta}$ & 2.11445 \AA$^{-1}$   \\
\hline
 \multicolumn{2}{|c|}{Derived constants:} \\
\hline
\multicolumn{2}{|l|}{equilibrium distance:\hspace{2.2cm} $R_e^X$= 4.06770(5) \AA} \\
\multicolumn{2}{|l|}{electronic term energy:\hspace{1.6cm}     $T_e^X$= -4217.815(10) \wn}\\
\hline
\end{tabular*}
\end{table}

\begin{table}
\fontsize{8pt}{13pt}\selectfont
\caption{Parameters of the analytic representation of the \astate state potential. The energy reference is the dissociation asymptote. Parameters with $^\ast$ are set for continuous extrapolation of the potential. }
\label{taba}
\begin{tabular*}{1.0\columnwidth}{@{\extracolsep{\fill}}|lr|}
\hline
   \multicolumn{2}{|c|}{$R < R_\mathrm{inn}=$ 4.956 \AA}    \\
\hline
   $A^\ast$ & -0.102755858$\times 10^{4}$ \wn \\
   $B^\ast$ & 0.595720017$\times 10^{6}$  \wn \AA $^{4}$ \\
\hline
   \multicolumn{2}{|c|}{$R_\mathrm{inn} \leq R \leq R_\mathrm{out}=$ 11.000 \AA}    \\
\hline
    $b$ &   $-0.41$              \\
    $R_\mathrm{m}$ & 5.90210000 \AA               \\
    $a_{0}$ &  -249.030467 \wn\\
    $a_{1}$ & -0.8249570532852936 \wn\\
    $a_{2}$ & 0.1696361682907917$\times 10^{4}$ \wn\\
    $a_{3}$ & 0.2195618553936388$\times 10^{4}$ \wn\\
    $a_{4}$ & -0.1830331320330776$\times 10^{5}$ \wn\\
    $a_{5}$ & -0.3577610246567447$\times 10^{6}$ \wn\\
    $a_{6}$ & 0.7605943179592050$\times 10^{6}$ \wn\\
    $a_{7}$ & 0.2321948211354688$\times 10^{8}$ \wn\\
    $a_{8}$ & -0.2969108359066523$\times 10^{8}$ \wn\\
    $a_{9}$ & -0.9066958465430735$\times 10^{9}$ \wn\\
   $a_{10}$ & 0.1193993210361681$\times 10^{10}$ \wn\\
   $a_{11}$ & 0.2271299134652605$\times 10^{11}$ \wn\\
   $a_{12}$ & -0.3822195877987927$\times 10^{11}$ \wn\\
   $a_{13}$ & -0.3733663448209457$\times 10^{12}$ \wn\\
   $a_{14}$ & 0.8258881465522462$\times 10^{12}$ \wn\\
   $a_{15}$ & 0.3983625994810529$\times 10^{13}$ \wn\\
   $a_{16}$ & -0.1170134952322679$\times 10^{14}$ \wn\\
   $a_{17}$ & -0.2572921814049761$\times 10^{14}$ \wn\\
   $a_{18}$ & 0.1083172601973219$\times 10^{15}$ \wn\\
   $a_{19}$ & 0.7316944602806914$\times 10^{14}$ \wn\\
   $a_{20}$ & -0.6401304321393455$\times 10^{15}$ \wn\\
   $a_{21}$ & 0.2090674461383090$\times 10^{15}$ \wn\\
   $a_{22}$ & 0.2236346785424132$\times 10^{16}$ \wn\\
   $a_{23}$ & -0.2555345293959939$\times 10^{16}$ \wn\\
   $a_{24}$ & -0.3556041967731776$\times 10^{16}$ \wn\\
   $a_{25}$ & 0.8321268009266810$\times 10^{16}$ \wn\\
   $a_{26}$ & -0.1364612053210692$\times 10^{16}$ \wn\\
   $a_{27}$ & -0.9030566262973000$\times 10^{16}$ \wn\\
   $a_{28}$ & 0.8950177388640356$\times 10^{16}$ \wn\\
   $a_{29}$ & -0.2723988913077922$\times 10^{16}$ \wn\\
\hline
   \multicolumn{2}{|c|}{$R_\mathrm{out} < R$}\\
\hline
  ${U_{\infty}}$ & 0.0 \wn    \\
 ${C_6}$ &    0.2072097$\times 10^{8}$ \wn\AA$^6$      \\
 ${C_{8}}$ &  0.6509487$\times 10^{9}$ \wn\AA$^8$   \\
 ${C_{10}}$ & 0.2575245$\times 10^{11}$ \wn\AA$^{10}$   \\
 ${A_{ex}}$ & -0.1469387$\times 10^{5}$ \wn\AA$^{-\gamma}$   \\
 ${\gamma}$ & 5.25669    \\
 ${\beta}$ & 2.11445 \AA$^{-1}$   \\
\hline
 \multicolumn{2}{|c|}{Derived constants:} \\
\hline
\multicolumn{2}{|l|}{equilibrium distance:\hspace{2.2cm} $R_e^a$= 5.9029(1) \AA} \\
\multicolumn{2}{|l|}{electronic term energy:\hspace{1.6cm}     $T_e^a$= -249.031(10) \wn}\\
\hline
\end{tabular*}
\end{table}

\section{Cold collisions}

As it was already mentioned in the introduction, Feshbach resonances for
$^{40}$K$^{87}$Rb were observed by several groups
\cite{Ferlaino:07, Ospelkaus}. In this section we apply the derived
potentials to calculate these Feshbach resonances and compare them
with observations. It is important to note that the spectroscopic
observations of the present study were done on different
isotopomers, thus application of the derived potentials for the
Feshbach resonances assumes the validity of the Born-Oppenheimer
approximation within the desired accuracy. We use for the
comparison the s-wave resonances observed by Ref. \cite{Ferlaino:07},
because these data represent the largest set and come from the
same lab, thus they should not show possible internal inconsistency
like calibration errors of the magnetic field. All calculated resonances were
found within 50 Gauss from the observed ones, this deviation is not surprisingly large because these
calculations are extrapolations out of the range of spectroscopic observations;
the largest outer turning point of observed levels is at 12.6 \AA
~for the singlet state and at 14.8 \AA ~for the triplet state
being 6.4 \wn and 2.3 \wn below the dissociation asymptote,
respectively. This indicates, that the spectroscopic observation
ends within the changeover to the long range behavior and the
quality of the extrapolation follows more from the reliability of
the theoretical estimates of the dispersion coefficients
\cite{Derevianko:01a, Porsev:03} than from our experimental data.

Thus the measured Feshbach resonances contain information on the
long range function. To include them in the total potential determination 
we set up a non-linear least squares fitting
code for the dispersion parameters, which contains the scattering
calculations as subroutine and searches for the maximum of the
elastic rate coefficient which corresponds to the observed
Feshbach resonance. The resonances have been observed by trap loss of
potassium, thus by an inelastic process, which will need three
body collisions. As very often assumed, the three body cross
section should be enhanced by the condition of a two body Feshbach
resonance. The calculations were done for a temperature of $1 \mu
K$ and use the accuracy of the magnetic field of 0.2 G, as stated in
Ref.~\cite{Ferlaino}, as uncertainty for each individual observation. In
a first step the resonances were fitted by varying the dispersion
terms $C_6$ and $C_8$, only. With the new total potentials asymptotic
levels of pure singlet (v = 97, 98, 99) and triplet (v=29, 30 31)
states were calculated and used as input for a new fit of the
single channel calculations. Error limits for these levels were
estimated from the fit of the resonances to be about 100 kHz.
In the new fit of the inner part of the potentials the coefficient
$C_{10}$ was chosen as an additional free parameter. With this result
a new iteration was started. Only two such iterations,
resonance fit followed by single channel fit, were needed for
convergence. The final results are given in table \ref{tabX} for
\Xstate~and in Table \ref{taba}~for \astate.

The ten s-wave Feshbach resonances could be fitted to a standard deviation
of $\sigma = 1.37$ and show on average a systematic deviation of
+0.05 G, i.e. the observations are at larger magnetic fields than the results from
the fit. This fit of the resonances is slightly better than the one
published in \cite{Ferlaino:07}: $\sigma = 1.66$ and mean
deviation +0.04 G, and has the additional advantage, that all
spectroscopic observations are equally well described as stated in
the previous section. All long range parameters are close to the
theoretical predictions \cite{Derevianko:01a, Porsev:03}, which
indicates that we have obtained a physically consistent model for the
two electronic states correlated to the atomic ground states 4s K
+5s Rb. It will be applied now to calculate an overview of
expected collision properties at low kinetic energy, the upper
limit of which is not easy to estimate because of the artificial
repulsive branches of the potentials (see eq. \ref{rep}).

Ferlaino et al. \cite{Ferlaino:07} give a list of singlet and
triplet scattering lengths for all possible isotope combinations of
KRb with natural abundance and calculate the scattering length of
the lowest hyperfine input channel of KRb. For the application of
mass scaling they used full potentials from other sources which
give for
$^{40}$K$^{87}$Rb as number of supported vibrational levels 98 for singlet and 32 for triplet. The
potentials derived in the present work support also 32 for
triplet, however 100 for singlet. Such possible variation was taken into account by the
error limits stated in \cite{Ferlaino}. Table \ref{scatlength}
compares our results with those from Ferlaino et al. The two columns, labeled as $a_{lowest}$, show scattering lengths of the
lowest Zeeman level in each case of the isotope pair, e.g. for
$^{39}$K + $^{85}$Rb f$_K$=1, m$_K$=1 and f$_{Rb}$=2, m$_{Rb}$=2.

\begin{table*}
\caption{Comparison of scattering lengths (in Bohr radius,
($a_0=0.5292 \times 10^{-10}$m).} \label{scatlength}
\begin{tabular}{r|rr|rr|rr} \hline
isotope & \multicolumn{2}{c|}{$a_{singlet}$} & \multicolumn{2}{c|}{$a_{triplet}$} & \multicolumn{2}{c}{ $a_{lowest}$} \\
& \cite{Ferlaino:07} & present & \cite{Ferlaino:07} & present & \cite{Ferlaino:07} & present  \\
\hline
    $39/85$ &   $26.5 (9)$    &   $ 33.4$ &   $63.0 (5)$ &   $ 63.9 $ &   $56.6(4)$ &   $ 58.0 $   \\
    $39/87$ &   $824^{+90}_{-70}$    &   $ 1868$ &   $35.9 (7)$ &   $ 35.90 $ &   $27.9(9)$ &   $ 28.36 $   \\
    $40/85$ &   $64.5 (6)$    &   $ 65.8$ &   $-28.4 (16)$ &   $ -28.55 $ &   $-25.3(16)$ &   $ -21.12 $   \\
    $40/87$ &   $-111 (5)$    &   $ -111.5$ &   $-215 (10)$ &   $ -215.6 $ &   $-185(7)$ &   $ -183.1 $   \\
    $41/85$ &   $106.0 (8)$    &   $ 103.1 $ &   $348 (10)$ &   $ 349.8 $ &   $283(6)$ &   $ 283.6 $   \\
    $41/87$ &   $14.0 (11)$ &   $ 7.06 $ &   $163.7 (16)$ &   $ 164.4 $ &   $1667^{+790}_{-406}$ &   $ 647.8 $   \\
\hline
\end{tabular}
\end{table*}

Certainly, for $^{40}$K$^{87}$Rb the results closely agree, but
for the other isotopomers significant differences, especially for
the singlet case appear, which reflects the different number of
supported vibrational levels within the respected potentials of
this state.

In the experiments by Ferlaino et al \cite{Ferlaino:07} and
Ospelkaus et al \cite{Ospelkaus} also p wave resonances and s wave
resonances coupled to d levels are reported. We checked if these
are also well described with the present potentials. Especially in
\cite{Ospelkaus} the splitting between $m_l=0$ and $m_l=\pm1$ of a
p wave is reported and it gives the opportunity to check if the
spin-spin or second order spin-orbit coupling for the triplet
state needs to be adjusted which both together can be described by the effective
operator \cite{Hirota}:

\begin{equation}
\label{SS}
 H_{SS}= \frac{2}{3} \lambda(3S_z^2-S^2).
\end{equation}

\noindent $\lambda$ is a function of internuclear 
separation R and typically is derived from spectroscopic data as an expectation value of a specific vibrational level. Here we use simple functional forms to incorporate the pure dipole-dipole contribution for the spin-spin interaction as a function $1/R^3$ and the spin-orbit part as an exponential function reflecting effectively an overlap integral of the electronic distribution of the two atoms:

\begin{equation}
\label{lambda}
 \lambda_{SS}= -\frac{3}{4} \alpha^2\left(\frac{1}{R^3}+a_\mathrm{SO}\exp{\left(-b(R-R_\mathrm{SO})\right)}\right)
\end{equation}

All quantities are given in atomic units, $\alpha$ is the fine
structure constant and $a_\mathrm{SO}$, b, and R$_\mathrm{SO}$ are
the model parameters for the second order spin-orbit contribution.
Since the data in hand cannot be sensitive to the actual function,
we selected values for b and $R_\mathrm{SO}$, which are comparable
to the example of Rb$_2$ \cite{Mies} (b= 0.7196 a$_0^{-1}$ and
R$_\mathrm{SO}$= 7.5 a$_0$) and fitted the parameter $a_{\mathrm
SO}$ which gives $a_\mathrm{SO}$= -0.013 a$_0^{-3}$ ($a_0=0.5292
\times 10^{-10}~m$). The result shows that the first part in eq.
\ref{lambda}  dominates in the internuclear separation interval $R
> 12 a_0 $ important for the bound levels of the triplet state
which determine the Feshbach resonances.

Taking this result into account, the four resonances reported in
\cite{Ferlaino:07, Ospelkaus} involving \textit{l} = 1 and 2 were
calculated. The \textit{l} = 1 resonances were found within the
experimental accuracy, and it is predicted that the resonance at
456G is split for m$_l$=0, $\pm$ 1 by about 0.1 G where the
component m$_l$=0 is higher in field than the others. This is
opposite to the case of the observed resonance at 515 G. The
\textit{l} = 2 resonances were found above the observed magnetic
field but not more than 0.7 G, the reason for this difference is presently
unclear. These resonances are very sharp, the calculated
rate constants of the corresponding elastic collisions have half
widths of less than 0.001 G, which depends on the magnitude of the
above mentioned SS coupling. No value was given in
\cite{Ferlaino:07} for their calculations of the widths.

\section{Conclusions}

An extensive dataset on the \astate\ state of KRb was collected
covering energy levels in a wide range of vibrational and
rotational quantum numbers in $^{39}$K$^{85}$Rb and
$^{39}$K$^{87}$Rb. The classical turning point of the last
observed level ($v'=25$, $J'=26$) lies at 14.6 \AA~on the fitted
rotationless potential energy curve. The long range parameters 
C$_6$, C$_8$, and C$_{10}$ deviate only slightly from the theoretical 
estimates by ref. \cite{Derevianko:01a, Porsev:03}, but we should remind 
the reader, that the present data set does not allow to determine independently 
all three parameters. Thus due to correlations between those of the data 
analysis, there is no contradiction to those of the theory.  But for 
modeling of cold collisions one should use the values reported in 
Tables~\ref{tabX} and \ref{taba} of the present work.

The complete determination of the potential curves allows for reliable 
extrapolation to the dissociation limit and thus their depths. 
Tables~\ref{tabX} and \ref{taba} give 4217.815(10) \wn~and 249.031(10) \wn~ for the respective dissociation energies.
Recently, Wang et al. \cite{Wang:07} reported about an independent determination 
of the dissociation energy of the singlet state using a photoassociation-depletion
process to obtain binding energies of loosely bound vibration levels, namely
$v=87$ and $89$. The dissociation energy was derived by adding level energies
with respect to the minimum of the potential curve as reported in Ref.~\cite{Amiot3:00}
by a near-dissociation expansion (NDE) approach. The final value 4217.822(3) \wn\ agrees
with ours within the given error limits. But for estimating the error limit Wang et al.
assume that the accuracy of level energy $v=87$ is equal to the standard deviation
of the NDE fit, 0.0015 \wn. Looking into the actual data set from Ref.~\cite{Amiot3:00} one finds that for 
$v=87$ only two high rotational states with $J$ = 50 and 52 were observed. Thus 
a fairly far extrapolation to the levels with $J=0$ is done, and one should certainly include 
the correlation of the NDE parameters for the error estimation. Additionally, we would remind the reader, 
that we corrected a misidentification of v=86 levels, due to overlapping lines, and this would effect the NDE analysis. Therefore, the good 
agreement between our present dissociation
energy and that reported by Wang et al. could be fortunate. As a second check we calculated 
the binding energies for $v=87$ and 89 from our results and found them larger by
0.026 \wn\ and 0.053 \wn, respectively. For $v=87$ the deviation is about 12 times
the error limit reported by Wang et al. We tried to incorporate the measured binding
energies in a new fit of the potentials including our spectroscopic results and
the Feshbach resonances from Ref.~\cite{Ferlaino:07},
but no convincing result was obtained up to now. The reason for the discrepancy 
is unclear for us, especially because we extended the dataset from Ref.~\cite{Amiot3:00}
just around $v=87$ with more $J$ levels and by measurements with $v=88$. Thus the energy
interval where the $v=87$ is located is well covered by our measurements which have 
an accuracy of typically 0.003 \wn.

By new spectroscopic data the potential curve of \Xstate could be
refined compared to the work by Ref. \cite{Amiot3:00}, and we were able
to link it to the potential curve of \astate with respect to
the asymptotic function and to the relative energy position of
both functions. This was the prerequisite to start the inclusion of
the observed Feshbach resonances \cite{Ferlaino} into the
modeling, which finally leads to a complete system for predicting
ultracold collision properties. Such prediction can be extended to
other isotopomers, as presented partly in Table \ref{scatlength}.
But it would be clearly desirable to get precise Feshbach resonances also for
different isotopomers in order to check the degree of validity of
the Born-Oppenheimer approximation.

Despite the combination of the spectroscopic and Feshbach data
there is a significant gap in the energy interval below the atomic
asymptote (about 3 \rcm) in which bound levels of low rotational
quantum numbers, preferable J or N =0, could be observed. Thus the
prediction in this region relies on the quality of the long range
parameters and measurements e.g. by two-color photo association
would be valuable. To reach such levels by Feshbach spectroscopy
is not promising because too high magnetic fields would be needed.

KRb is very often discussed for applications in studies of dipolar cold
gases or even quantum computing. To transfer Feshbach molecules to
deeply bound levels which are stable and would have large electric
dipole moments, spectroscopic data on excited states will be of
importance to predict efficient processes. Such data are partly existing
from other authors and are in hand also from our own spectroscopic work, which
will be analyzed joining the available data. Of particular interest
in KRb is the HFS in the B$^1\Pi$ state which is supposed to cause
the ``irregular'' hyperfine splitting the the transitions to the
\astate\ state.

\section{Acknowledgments}

We acknowledge gratefully C. Amiot for providing us his total experimental data in electronic form. 
The work is supported by DFG through SFB 407 and GRK 665. O.D.,
M.T. and R.F. acknowledge the support by the NATO SfP 978029
Optical Field Mapping grant and by the Latvian Science Council
grant No. 04.1308. \ O.D. acknowledges the support from the
European Social Fund and A.P. - a partial support from the
Bulgarian National Science Fund Grants MUF 1506/05 and VUF 202/06.


\begin{thebibliography}{10}
\bibitem{hundb}
{the appropriate rotational quantum number for the \astate\ state is N in the Hund's case b}
\end{thebibliography}

\end{document}